\documentclass[journal]{IEEEtran}
\usepackage{graphicx}
\usepackage{dcolumn}
\usepackage{bm}
\usepackage{caption}
\usepackage{comment}
\usepackage{subcaption}
\usepackage{xcolor}
\usepackage{physics}
\usepackage{ulem}
\usepackage{placeins}
\usepackage{booktabs}
\usepackage{tabularx}
\usepackage{siunitx}
\usepackage{todonotes}

 \usepackage{algpseudocode}

\usepackage{algorithm}

\begin{document}

\title{Higher Order Charge Conserving Electromagnetic Finite Element Particle in Cell Method}

\author{Z.~D.~Crawford,~\IEEEmembership{Student~Member,~IEEE,}
        O.~H.~Ramachandran,~\IEEEmembership{Student~Member,~IEEE,}
        S.~O'Connor,~\IEEEmembership{Student~Member,~IEEE,}
        J.~Luginsland,~\IEEEmembership{Fellow,~IEEE,}
        and~B.~Shanker,~\IEEEmembership{Fellow,~IEEE}
        
        \thanks{  Z. D. Crawford, S. O'Connor, B. Shanker are with the Department
of Electrical and Computer Engineering, Michigan State University, East Lansing,
MI, 48824.\protect\\
 J. Luginsland is with AFRL/Air Force Office of Scientific Research, Arlington, VA 22201.
E-mail: crawf326@msu.edu}
}

\markboth{IEEE Transactions on Plasma Sciences,~Vol.~14, No.~8, Feb~2020}%
{Shell \MakeLowercase{\textit{et al.}}: Bare Demo of IEEEtran.cls for IEEE Journals}

\maketitle
\begin{abstract}
Until recently, electromagnetic finite element PIC (EM-FEMPIC) methods that demonstrated charge conservation used explicit field solvers. It is only recently, that a series of papers developed the mathematics necessary for charge conservation within an implicit field solve and demonstrated for a number of examples. This permits using time steps sizes that are necessary to capture the physics as opposed to being restricted to those constrained by geometry. One aspect that is missing is higher order basis functions to represent both fields and particles. Higher order basis can be particularly helpful in effectively capturing complex field layouts with fewer degrees of freedom. Developing a framework for higher order EM-FEMPIC that maintains stability, improves accuracy, and conserves charge is the principal goal of this paper. A number of results are presented that attest to its efficacy.  

\end{abstract}


\IEEEpeerreviewmaketitle

\section{Introduction}\label{sec:introduction}

Modeling novel beam-wave interaction devices, such as accelerators, vacuum electronics, and directed energy devices \cite{peterkin2002virtual,shin2011,Cooke2013} relies on robust numerical tools capable of self-consistent analysis of the  interaction of a plasma with electromagnetic fields. This is typically done using an electromagnetic particle-in-cell (EM-PIC) method to evolve a given plasma distribution in time and space \cite{birdsall2018plasma}.
It consists of a method that discretizes both the Newton's equations of motion and a Maxwell's field solver. The coupling between the two is effected through the Lorentz force due to the electric field and magnetic flux density. Given the range of applications, there has been extensive interest in developing PIC solvers; a majority of those used in field are based on finite difference time domain (FDTD) methods \cite{meierbachtol2015conformal}. The simplicity of the formulation, ease of particle position updates, and readily available parallelization algorithms make this approach an attractive workhorse for PIC. In what follows, we will use a concatenation of abbreviations to denote regime and method to solve PIC. For instance, EM-FDTDPIC denotes an electromagnetic PIC using FDTD.

While EM-FDTDPIC has a number of advantages, there has been significant recent effort to explore the advantages offered by finite element methods to solving PIC problems. To a large part, this is driven by success of this methodology in microwave and millimeter frequency regimes wherein the success of this method has been demonstrated in analysis and design of complex topologies and electrically large objects. The foray of FEM into PIC is not without challenges, the principal of which is charge conservation. To understand this, note that in evolving the fields, we only solve the two curl equations (Faraday's and Ampere's laws) and need a framework wherein Gauss' law are satisfied as well. This implies that discretization in space \emph{and} time should be such that these laws are satisfied. Ref. \cite{crawford2021rubrics} rigorously develops the conditions that should be satisfied, and demonstrates how current EM-PIC formulations satisfy these conditions. For instance, the spatial and temporal basis sets used in an explicit FDTD time-stepping scheme, together with an appropriate integration of the path, satisfies these constraints. Developing such a method that was efficient for FEM was a long standing challenge. 

This was rigorously solved recently; see pioneering papers by \cite{squire2012geometric,pinto2014charge,moon2015exact,o2021set}. The methods introduced here were  based on explicit updates of field solution and particle position, and on the proper representation of quantities on the underlying discrete mesh. In the same vein, a Poisson bracket approach that utilizes Whitney forms defined by B-spline FEM formulations \cite{buffa2010isogeometric} to define a structure-preserving EM-PIC scheme \cite{kraus2017gempic,jianyuan2018structure}, with several of these methods using higher order basis sets. Note, in manner akin to FDTD, one solves all of Maxwell's equations. In an explicit setting, this approach avoids exciting null spaces (and corruption of Gauss' law due to these null spaces) \cite{campos2021variational,peifeng2021discovering,perse2021geometric}.

But restriction of field solve to an explicit field update has challenges; it is only conditionally stable and the smallest time step is governed by the finest feature in the model and \emph{not} the physics. Overcoming this bottleneck has a well known remedy. Indeed, Newmark-Beta methods are unconditionally stable and the constraint on time comes from the physics that one needs to captures. But implicit field solve implies the need to rethink PIC solves such that the involution Gauss' laws are satisfied. A problem that is unstated is that implicit solves introduce a null space; for Maxwell solvers this null space is of the form $\nabla \phi (\vb{r})$, and for the wave equation this null space is of the form $t \nabla \phi (\vb{r})$. It is apparent that the null space will corrupt the satisfaction of Gauss' laws in addition to other challenges. This problem was solved recently \cite{o2021time,o2021quasi}. Specifically, imposition of Coulomb Gauge using a quasi-Helmholtz decomposition (in simply connected systems) in \cite{o2021quasi} enables satisfying Gauss' law to machine precision for \emph{both} the Maxwell solver and the wave equations. This implies that the infrastructure that is already in place to solve the vector wave equation can readily used for PIC analysis. 

As we build this line of progress, the next ingredient that is missing is higher order basis sets for field, current and particle representation within a PIC framework. Hierarchical and interpolatory basis functions are known for FEM field solvers \cite{hiptmair2001higher,graglia1997higher,graglia2012hierarchical}.
For smoothly varying geometries, higher order bases provide more accurate fields while utilizing fewer degrees of freedom.
Furthermore, these higher order formulations satisfy the relationships of the \emph{de-Rahm} complex \cite{deschamps81,boss88,arnold10,warnick14}. Developing a higher order EM-FEMPIC framework will be the key contribution of this paper. Specifically, we will present an unconditionally stable, exact current mapping FEM EM-PIC scheme that uses higher order basis functions on tetrahedral meshes.

The rest of this paper is organized as follows:  In Section \ref{sec:Prelim} provide a brief problem statement. Next, in Section \ref{sec:Analysis}, we define the spatial and temporal discretization of the problem. Section \ref{sec:mapping} describes the current mapping scheme used to conserve charge regardless of the time marching scheme.  In Section \ref{sec:results}, we present results that demonstrate the use of the higher order FEM-PIC scheme. Finally, we conclude this paper in Section \ref{sec:conclusions} outlining future directions of research. 

\section{Problem Statement \label{sec:Prelim}}

Consider a region of free space $\Omega$ containing charged species. The boundary of the $\Omega$ is denoted by $\partial \Omega$.  For simplicity we consider only a single species. 
The permittivity and permeability of free space are denoted as $\varepsilon_0$ and $\mu_0$,  and the speed of light denoted using $c = 1/\sqrt{\mu_0\varepsilon_0}$. There also exists a time-varying electromagnetic field due to moving charges and potentially an impressed electromagnetic field. The distribution of the charge species is  represented by a phase space distribution function (PSDF) $f(t,\vb{r},\vb{v})$ that satisfies the Vlasov equation 
\begin{align} \label{eq:vlasov}
  \partial_t f(t,\vb{r},\vb{v})  + \vb{v} \cdot \nabla f(t,\vb{r},\vb{v}) + \\ \frac{q}{m} [\vb{E}(t,\vb{r}) + \vb{v} \times \vb{B}(t,\vb{r})] \cdot \nabla_v f(t,\vb{r},\vb{v}) = 0. \nonumber
\end{align}

\section{Overview of Discrete Solutions \label{sec:Analysis}}

In what follows, we follow the usual path of representing the moments of distribution function via the charge and current density as, $\rho(t,\vb{r}) = q \int_\Omega f(t,\vb{r},\vb{v}) d\vb{v}$ and $\vb{J}(t,\vb{r}) = q\int_\Omega \vb{v}f(t,\vb{r},\vb{v})d\vb{v}$. Using a particle approximation with $N_p$ shape functions $S(\vb{r})$, one obtains
\begin{subequations}
\begin{equation}
    \rho(t,\vb{r}) = q\sum_{p=1}^{N_p} S(\vb{r}-\vb{r}_p(t))
\end{equation}
\begin{equation}
    \vb{J}(t,\vb{r}) = q\sum_{p=1}^{N_p} \vb{v}(t)S(\vb{r}-\vb{r}_p(t))
\end{equation}
\end{subequations}
where $\vb{r}_p(t)$ and $\vb{v}_p(t)$ are the position and velocity of particle $p$. In this work the shape functions are chosen to be Dirac delta functions, though generalization to other shape functions is possible \cite{crawford2021rubrics}.  The particular choice shape function is immaterial to the results of this paper. What we seek is the self-consistent evolution of the charge and current densities due to electromagnetic field resulting from the equations of motion of particles. This calls for a self consistent solution to Maxwell's equation and equations of motion. 

For completeness,  electromagnetic fields satisfy Maxwell's curl equations 
\begin{subequations}\label{eq:maxwell_cont}
\begin{equation}\label{eq:faraday}
        - \frac{\partial \vb{B}(t,\vb{r})}{\partial t} = \curl \vb{E}(t,\vb{r})
\end{equation}
\begin{equation}\label{eq:ampere}
      \frac{\partial \vb{D}(t,\vb{r})}{\partial t} = \curl \vb{H}(t,\vb{r}) - \vb{J}(t,\vb{r})
\end{equation}
\end{subequations}
and Gauss' laws
\begin{subequations}\label{eq:gauss_cont}
\begin{equation}\label{eq:gaussB}
    \div \vb{B}(t,\vb{r}) = 0
\end{equation}
\begin{equation}\label{eq:gaussE}
    \div \vb{D}(t,\vb{r}) = \rho(t,\vb{r}).
\end{equation}
\end{subequations}
where $\vb{E}(t,\vb{r})$, $\vb{D}(t,\vb{r})$, $\vb{H}(t,\vb{r})$ and $\vb{B}(t,\vb{r})$ are the electric field, electric flux density, magnetic field and magnetic flux density, respectively. The fields are  subject to boundary conditions which are either Dirichlet, Neumann, or impedance boundary conditions on $\partial\Omega_D$, $\partial\Omega_N$, or $\partial\Omega_I$ which bound the domain as
\begin{subequations}\label{eq:bceq}
\begin{equation}\label{eq:dirichlet}
\hat{n}\times \mathbf{E}(t,\mathbf{r}) = \mathbf{\Psi}_D(t,\mathbf{r})\;\;\text{on}\;\partial\Omega_D,
\end{equation}
\begin{equation}\label{eq:dirichlet}
\hat{n}\times \mathbf{H}(t,\mathbf{r}) = \mathbf{\Psi}_N(t,\mathbf{r})\;\;\text{on}\;\partial\Omega_N,
\end{equation}
\begin{equation}\label{eq:impedanceBC}
\hat{n}\times \frac{\mathbf{B}(t,\mathbf{r})}{\mu} - Y\hat{n}\times\hat{n}\times \mathbf{E}(t,\mathbf{r}) = \mathbf{\Psi}_I(t,\mathbf{r})\;\;\text{on}\;\partial\Omega_I.
\end{equation}
\end{subequations}
As is to be expected, free space consititutive relations $\vb{D}(t,\vb{r}) = \varepsilon_0\vb{E}(t,\mathbf{r})$ and $ \vb{B}(t,\mathbf{r}) = \mu_0\vb{H}(t,\vb{r})$ hold. 
The particle position of the sources are evolved using Newton's equations of motion and Lorentz force, viz., $\vb{F} (t, \vb{r})  = q (t, \vb{r}) (\vb{E} (t, \vb{r}) + \vb{v} (t, \vb{r}) \times \vb{B}(t, \vb{r}) )$. 
The simulation follows the usual PIC cycle: particles are mapped to a discretized space to solve for the electric field and magnetic flux density, which are in turn used to push the particles, defining a new current and particle positions, and so on. Note, 
although the PSDF is sampled with particles, the total description of the electromagnetic problem is continuous, and must be discretized in space and time. 

\subsection{Discretization in Space}

\begin{figure}
\centering
\includegraphics[scale=.3]{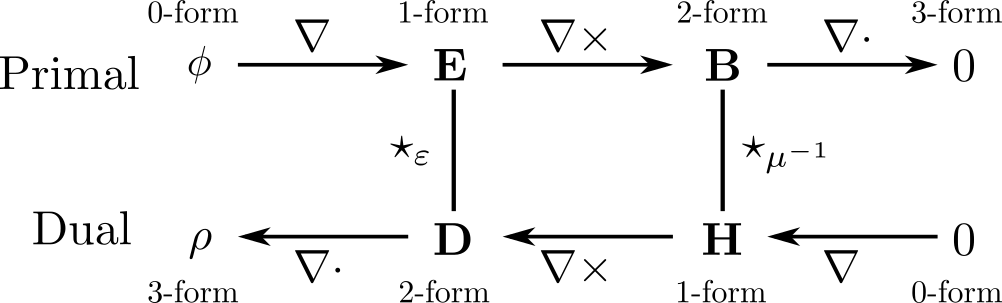}
\caption{ Relationship of fields and flux densities with respect to the \emph{de-Rahm} complex. }
\label{fig:primdual}
\end{figure}

Assume that the domain $\Omega$ is represented using a collection of finite elements $\mathcal{K} = \left \{ \mathcal{N}, \mathcal{E}, \mathcal{F}, \mathcal{T} \right \}$ defined using $N_n$ nodes, $N_e$ edges, $N_f$ faces and $N_t$ tetrahedron. Each tetrahedron contains basis functions to represent fields, flux densities, and sources that follow the \emph{de-Rham} sequence as seen in Fig. \ref{fig:primdual}\cite{graglia1997higher,graglia2012hierarchical}. 
This sequence preserves the differential relations between the quantities of interest, such that the curl of a field is a flux density, the divergence of a flux density is a charge.  The Hodge star operator, $\star$,  maps a field to a flux density on a dual mesh.
It is well known that Whitney basis functions can be used to represent the electric field and magnetic flux density \cite{monk2003finite,jin2015finite,pinto2014charge}.

For a $k$th order interpolatory basis functions, the electric field is represented using $N_1$ higher order Whitney edge basis functions, $\vb{E}(t,\vb{r}) = \sum_{i=1}^{N_1} e_i(t) \vb{W}^{(1)}_{i}(\vb{r})$, where there are $6(k+1)$ degrees of freedom associated with edges, $4k(k+1)$ degrees of freedom associated with faces, and $k(k^2-1)/2$ associated with the cell volumes. The magnetic flux density is represented using $N_2$ Whitney face basis function,  $\vb{B}(t,\vb{r}) = \sum_{i=1}^{N_2} b_i(t) \vb{W}^{(2)}_{i}(\vb{r})$ where there are $(k+1)(k+2)/2$ degrees of freedom associated with faces and $k(k+1)(k+2)/2$ associated with the cell volumes.
A complete description of these basis functions is provided in the Appendix.

We have chosen to define the problem such that Faraday's law in \eqref{eq:faraday} (and the corresponding quantities) is defined on the primal grid.  
Therefore, Ampere's law \eqref{eq:ampere}, $\vb{H}(t,\vb{r})$, $\vb{D}(t,\vb{r})$, \emph{as well as} $\vb{J}(t,\vb{r}),\and \rho(t,\vb{r})$ are defined in the dual space on the corresponding dual grid.
This means that while the electric field and magnetic flux densities can be directly represented using a Whitney basis on the primal mesh, the sources $\rho(t,\vb{r})$, $\vb{J}(t,\vb{r})$ cannot.  On structured grids it is straightforward to define dual basis function spaces to represent dual quantities, however, on an unstructured FEM mesh the dual fields are only indirectly accessible via Hodge operators.
Therefore the source distribution $\rho(t,\vb{r})$ cannot be directly represented, and is instead measured with higher order nodal basis functions on the primal mesh, which are defined in the Appendix. The current density $\vb{J}(t,\vb{r})$, which lies in the same space as $\vb{D}(t,\vb{r})$, is measured by the the electric field basis, the higher order Whitney edge functions.

Before we proceed with prescribing the discrete framework, consider an auxillary function  
\begin{align}\label{eq:G}
    \vb{G}(t,\vb{r}) = \int_0^{t} \vb{J}(\tau,\vb{r}) d\tau
\end{align}
such that Ampere's law is rewritten as
\begin{equation}\label{eq:ampere2}
      \frac{\partial \vb{D}(t,\vb{r})}{\partial t} = \curl \vb{H}(t,\vb{r}) - \frac{\partial \vb{G}(t,\vb{r})}{\partial t}.
\end{equation}
Using \eqref{eq:ampere2} and Faraday's law and spatial basis functions defined earlier,  one may write the discrete system as 
\begin{align}\label{eq:maxwell_semi}
\begin{split}
    \underbrace{\mqty[ [\star_{\mu^{-1}}] & 0 \\ 0 &[\star_{\epsilon_0}]]}_{\Bar{\Bar{M}}} &\mqty[\partial_t \bar{B}(t)\\ \partial_t \bar{E}(t) ] \\
    &\;\;+ \underbrace{\mqty[0&  [\tilde{\nabla}\times] \\ -[\tilde{\nabla}\times]^T & [\star_I] ]}_{\Bar{\Bar{S}}} \mqty[\bar{B}(t) \\\bar{E}(t) ] = \underbrace{\mqty[0\\ -\frac{\partial_t\bar{G}(t)}{\epsilon} ]}_{\bar{\bar{F}}}
\end{split}
\end{align}
where the degree of freedom vectors $\bar{E}(t) = [e_1(t),e_2(t),\dots,e_{N_e}(t)]$, $\bar{B}(t) = [b_1(t), b_2(t),\dots,b_{N_f}(t)]$, and $\bar{G}(t)=[g_1(t),g_2(t),...g_{N_e}(t)]$ with $g_i (t) = \langle \vb{W}_i^{(1)} (\vb{r}), \vb{G} (t, \vb{r}) \rangle $.  
The coupled system matrix is composed of discrete Hodge matrix operators
\begin{equation}
        [\star_\epsilon]_{i,j} = \langle \vb{W}^{(1)}_i(\vb{r}),\varepsilon\cdot\vb{W}^{(1)}_j(\vb{r}) \rangle
\end{equation}
\begin{equation}
        [\star_{\mu^{-1}}]_{i,j} = \langle \vb{W}^{(2)}_i(\vb{r}),\mu^{-1}\cdot\vb{W}^{(2)}_j(\vb{r})\rangle,
\end{equation}
the surface impedance matrix
\begin{equation}
        [\star_I]_{i,j} = \langle \hat{n}_i\cross\vb{W}^{(1)}_i(\vb{r}),\mu^{-1}\hat{n}_j\cross\vb{W}^{(1)}_j(\vb{r})\rangle_{\partial\Omega}
\end{equation}
and a discrete curl matrix
\begin{equation}
        [\tilde{\grad}\times]_{i,j} = \langle \vb{W}^{(2)}_i , \curl \vb{W}^{(1)}_j(\vb{r})\rangle.
\end{equation}
The operator $\langle\cdot,\cdot\rangle$ and $\langle\cdot,\cdot\rangle_{\partial\Omega}$ define a volume and surface integral, respectively, over the support of the basis functions, which is either a tetrahedron or face.
The discrete curl matrix $[\tilde{\grad}\times]$ includes the metric information, unlike the definition usually obtained through discrete exterior calculus $[\curl]$, which has entries of only 0,+1, or -1. For lowest order $k=1$, this definition can be obtained as 
\begin{equation}
    [\curl]=[\star_{\mu^{-1}}]^{-1}[M_c];
\end{equation}
however, the simplicity of defining the matrix by inspection is lost for $k>1$.

\subsection{Evolution of Particle Path and Current Mapping}\label{sec:mapping}

The fields are evolved in time using an unconditionally stable Newmark-beta time marching scheme \cite{zienkiewicz1977new,crawford2020unconditionally}.
This allows larger time steps than would be afforded by a leapfrog method. In this framework, the current mapping (or evolution of charge) has to be consistent with that used for evolution of fields. Unfortunately, a naive approach to incorporate the particle current as the forcing function will violate conservation of charge. The method presented in \cite{o2021time} is agnostic to any time stepping method used for a field solve, and overcomes this bottleneck. In this paper, a similar method is used, but adapted to a higher order basis function in space. 
Using the definition of the particle current density in \eqref{eq:G}, the Newmark-Beta time marching scheme is defined as
 \begin{equation}\label{eq:maxnmb}
 \begin{split}
     (\gamma\bar{\bar{M}} +\beta\Delta_t\bar{\bar{S}})\bar{X}^{n+1} &-\gamma\Delta_t\bar{\bar{S}}\bar{X}^n
     +(\gamma\bar{\bar{M}} +\beta\Delta_t\bar{\bar{S}}_M)\bar{X}^{n-1}\\ &+\gamma\Delta_t\bar{\bar{F}}^{n+1}     + \gamma\Delta_t\bar{\bar{F}}^{n-1}=0
 \end{split}
 \end{equation}
 where $\bar{X}^m = [\bar{B}^T(t_m)\; \bar{E}^T(t_m)]$ and $\bar{\bar{F}}^m = [ 0\;\; -\varepsilon^{-1}\bar{G}^T(t_m)]$.
The degree of freedom vector $\bar{G} = [g_1(t),g_2(t),\cdots,g_{N_e}(t)]$ with $g_i(t) = \langle \vb{W}^{(1)}_i,\vb{G}(t,\vb{r})\rangle$. The parameters $\gamma$ and $\beta$ are chosen to be $\gamma=.5$ and $\beta=.25$.
To define the forcing function $\tilde{F}$, it is necessary to use an integration rule appropriate for the product of the higher order edge basis function and the particle path, which may also be a higher order polynomial. A key point that should be noted is the de-linking of the time stepping algorithm used for particle push and field updates. Consistency is ensured by proper inclusion in the right hand side of \eqref{eq:maxwell_semi}. Note, that this assumes non-relavistic motion. In this work, a fourth order Adams-Bashforth push is used, making the particle path a fourth order Lagrange polynomial. 

\subsection{Satisfaction of Gauss's Magnetic Law}

It is well known that the Newmark-Beta solution to Maxwell's equations suffers from a null space that \emph{does not} grow in time. The amplitude of the excited null space corresponds to the accuracy of the solution at every time step. Despite this, as is evident in \cite{o2021quasi}, under a number of conditions, this null space does not corrupt the overall solution. Our challenge is when the impressed fields are strong. We have shown in \cite{o2021quasi}, that imposing the Coulomb gauge ensures that even if null spaces are generated they do not corrupt the satisfaction of Gauss' laws. This was done using topological approach. Here, our goal is to explore an alternative approach, targeted at ensuring that impressed magnetic flux densities are divergence free. For example, as we seek to examine particle motion due to impressed magnetic lenses, we want to ensure the discrete representation of impressed magnetic flux density is divergence free when represented using Whitney basis. To that end, consider reconstructing an impressed magnetic flux density $\vb{B}^{i}(t,\vb{r})$ which satisfies \eqref{eq:gaussB}.
The usual approach to obtain the coefficients to approximate $\vb{B}^{i}(t,\vb{r})$ would be to use Galerkin testing with the divergence-conforming basis set
\begin{equation}
    [\star_{\mu^{-1}}]\vb{b} = \vb{f}
\end{equation}
where $\vb{f}_i = \langle \vb{W}^{(2)}_i(\vb{r}), \vb{B}^{i}(t,\vb{r})\rangle$.
However, this construction of $\tilde{\vb{B}}^{i}(t,\vb{r})$ will not satisfy Gauss's law unless $\tilde{\vb{B}}^{i}(t,\vb{r}) = \vb{B}^{i}(t,\vb{r})$.
This is accomplished by solving an optimization problem where Gauss's magnetic law is the constraint.

\begin{equation}\label{eq:conB}
\begin{bmatrix}
[\star_{\mu^{-1}}] &[\div]^T\\
[\div] & 0
\end{bmatrix}
\begin{bmatrix}
\vb{b}\\\lambda
\end{bmatrix}
=
\begin{bmatrix}
\mathcal{B}\\0
\end{bmatrix}
\end{equation}
For the lowest order spatial basis functions, the discrete divergence operator $[\div]$ can be written by inspection.
The discrete divergence operator for higher orders, like the discrete curl operator, cannot be written as easily.
It can be written as
\begin{equation}
    [\div] = [\star_3]^{-1}[\tilde{\nabla}\cdot]
\end{equation}
where 
\begin{equation}
    [\star_3]_{ij}=\langle W^{(3)}_i(\vb{r}),W^{(3)}_j(\vb{r})\rangle
\end{equation}
and
\begin{equation}
    [\tilde{\grad}\cdot]_{ij}=\langle W^{(3)}_i(\vb{r}),\div \vb{W}^{(2)}_j(\vb{r})\rangle
\end{equation}
with $W^{(3)}(\vb{r})$ as the higher  order  volumetric  basis function, which is defined in the Appendix.
This allows the reconstructed field, regardless of the error in the representation of the original function, to still satisfy \eqref{eq:gaussB}.

Consider data provided in Table \ref{tab:divblin} and \ref{tab:divbsin}. In this test represented in the tables, a divergence free function was reconstructed in a volume 50 cm $\times$ 15 cm $\times$ 20 cm, for several orders of spatial basis functions. In Table \ref{tab:divblin}, the function $\vb{B}(\vb{r}) = y\hat{x} + x\hat{y}$ can be reconstructed exactly with second order and higher basis functions. Therefore, once the function was modeled correctly to machine precision, the divergence free nature of the reconstructed field is seen in both the constrained and non-constrained formulation. However, in Table \ref{tab:divbsin}, the function used is $\vb{B}(\vb{r}) = \sin(y)\hat{x} + \cos(x)\hat{y})$, which cannot be represented exactly by a finite set of polynomials. Despite having similar accuracy, only the constrained formulation satisfies \eqref{eq:gaussB}. It is evident that the formulation has the desired properties with respect to representation of fields. As alluded to earlier, we will use this only for representing only the impressed field. A topological approach, akin to \cite{o2021quasi}, is being developed and will be presented in a later paper. 

\begin{table}
\addtolength{\tabcolsep}{-4pt}
\centering
\caption{Comparison of Error in Field Reconstruction and Gauss' Magnetic Law for Constrained and Non-constrained Linear Field} \label{tab:divblin}  
\begin{tabular}{|c | c c | c c |}
\hline
& \multicolumn{2}{c|}{Non-constrained} &\multicolumn{2}{c|}{Constrained} \\
  $m$ &  error in B & error in  $\div\vb{B}$ & error in B & error in $\div\vb{B}$\\
\hline
 1 & $1.0604\times 10^{-1}$  & $2.9681\times 10^{-5}$  & $1.1367\times 10^{-1}$  & $4.5062\times 10^{-19}$ \\
 2 & $1.1049\times 10^{-15}$ & $6.1606\times 10^{-19}$ & $1.2201\times 10^{-15}$ & $4.4173\times 10^{-19}$ \\
 3 & $3.8299\times 10^{-15}$ & $1.9566\times 10^{-18}$ & $9.8539\times 10^{-15}$ & $3.9481\times 10^{-18}$ \\
\hline
\end{tabular}
\end{table}

\begin{table}
\addtolength{\tabcolsep}{-3pt}
\centering
\caption{Comparison of Error in Field Reconstruction and Gauss' Magnetic Law for Constrained and Non-constrained Sinusoidal Field} \label{tab:divbsin}  
\begin{tabular}{|c | c c | c c |}
\hline
& \multicolumn{2}{c|}{Non-constrained} &\multicolumn{2}{c|}{Constrained} \\
  $m$ &  error in B & error in $\div\vb{B}$ & error in B & error in $\div\vb{B}$\\
\hline
 1 & $1.3198\times 10^{-2}$ & $1.6310\times 10^{-5}$ & $1.5048\times 10^{-2}$ & $1.1404\times 10^{-18}$ \\
 2 & $6.1965\times 10^{-4}$ & $3.1300\times 10^{-7}$ & $5.8211\times 10^{-4}$ & $1.2061\times 10^{-18}$ \\
 3 & $6.3122\times 10^{-6}$ & $2.7822\times 10^{-9}$ & $7.0793\times 10^{-6}$ & $1.0762\times 10^{-17}$ \\
\hline
\end{tabular}
\end{table}

\section{Results\label{sec:results}}

In this Section, we present several numerical tests using higher order FEM-PIC. Particle free results are provided to demonstrate correctness of our implementation. We demonstrate that the higher order basis functions presented satisfies the continuity equation and Gauss' law for a number of cases using the stated particle mapping scheme. 

\subsection{Cost of Higher Order Representation}

\begin{figure}
    \centering
    \includegraphics[draft=false,scale=0.55]{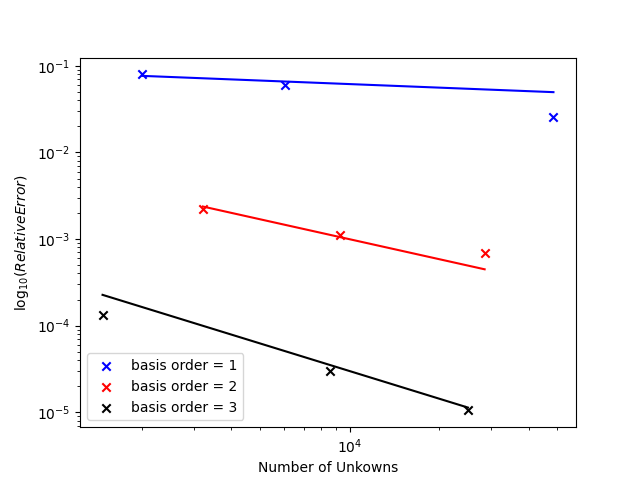}
    \caption{Relative Error in $\vb{E}_{t,\vb{r}}$ for field propagating through free space. }
    \label{fig:covnErr}
\end{figure}

First,  we present the error in computed fields with respect to number of unknowns and basis function order.  The relative error in a field propagating through a region of free space is shown in Fig. \ref{fig:covnErr}. The region is \SI{.25}{\metre} $\times$ \SI{.2}{\metre} $\times$ \SI{.5}{\metre} in size. The normally incident electric field is defined as 
\begin{equation}\label{eq:excite}
\bar{E}(\bar{r},t) = \hat{y}\cos(2\pi f_0 \tau)e^{-(\tau  - 8\sigma)^2/2\sigma^2}\, \text{({V/m})},
\end{equation}
where where $\tau = t-\bar{r}\cdot\hat{z}/c$, $\sigma = 3/[2(f_\text{max} - f_0)]$ with the center $f_0= 100 MHz$ and maximum frequency $f_\text{max}=195 MHz$.
 The relative error is defined as 
\begin{equation}\label{eq:rerr}
    \text{relative error} = \frac{||\vb{E}_i^{(r)}-\vb{E}_i^{(a)}||_2}{||\vb{E}^{(a)}||_2}
\end{equation}.
As is evident from Fig. \ref{fig:covnErr}, it take several orders of magnitude more unknowns for a first order basis function to reach the same level of accuracy as a second order basis function. The trade off is that the condition number of the system increases by roughly an order of magnitude as the order increases which effects the rate at which an iterative solver will converge.
Therefore, consideration can be taken in balancing the size of the problem and simulation time for a given error.  

\subsection{Higher Order Particle Motion}

\begin{figure}
    \centering
    \includegraphics[draft=false,scale=0.55]{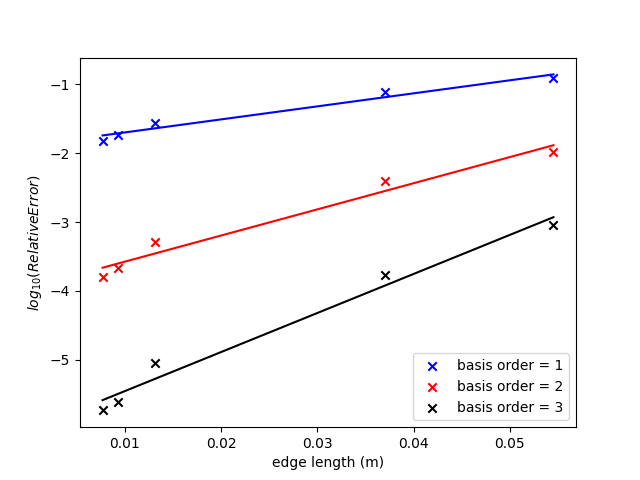}
    \caption{Relative Error in $\vb{E}_{\rho}$ for orbital motion}
    \label{fig:CycRfield}
\end{figure}

\begin{figure}
    \centering
    \includegraphics[draft=false,scale=0.55]{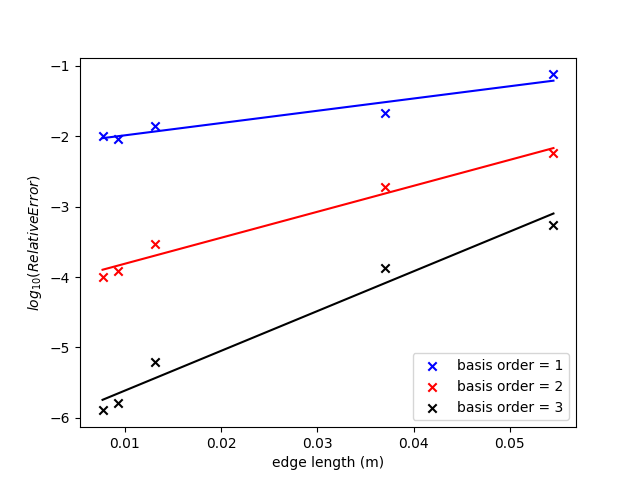}
    \caption{Relative Error in $\vb{E}_z$ for orbital motion.}
    \label{fig:CycZfield}
\end{figure}

The first example with particles is the orbit of a single particle around a nucleus. This test demonstrates when higher order bases give more accurate particle trajectories. At a certain distance from nucleus, a particle with initial velocity perpendicular to the radial electric field will result in a circular orbit in a plane. 
In this test, a particle with initial velocity $\vb{v} = 3\times10^6\hat{\phi}$m/s is set .25m from the centroid of a cylindrical ring geometry. 
The geometry has an inner radius of .2m and outer radius of .3m.
Five meshes were generated to compare refinement in the edge length to basis function order. 
The electric field due to a nucleus with a charge of $Q_n = 1.423 nC$ is reconstructed using the higher order interpolatory Whitney edge basis set and used to push the electron in the geometry.
The experiment was run for 6000 time steps with $\Delta_t = $0.4ns, which corresponds to approximately 4 cycles.
The effect of the higher order basis functions can be seen in \ref{fig:CycRfield} and \ref{fig:CycZfield} where the relative error is defined in \eqref{eq:rerr}.

\begin{figure}
    \centering
    \includegraphics[draft=false,scale=0.55]{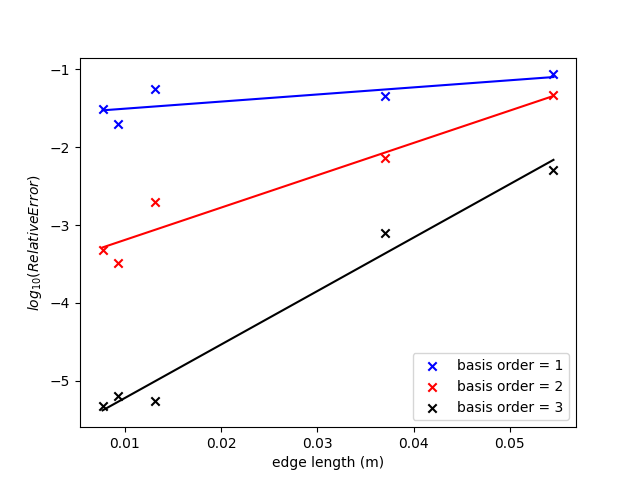}
    \caption{Relative Error in $\vb{r}$ for orbital motion.}
    \label{fig:CycMotion}
\end{figure}

For both the $\hat{z}$ and $\hat{\rho}$ components of the electric field, the error in the fields converges.
The improvement of the field error translates to an improvement of the particle trajectory.
In \ref{fig:CycMotion}, the relative error is shown for the $\hat{z}$ component of the particle position.
A key takeaway is that using a higher order basis function leads to more accurate particle trajectories than simply refining the mesh.

\subsection{Plasma Ball}
 In this example, we simulate an adiabatic expansion of a plasma ball. This example has both approximate analytic solutions \cite{kovalev2003analytic} as well as experimental data \cite{laha2007experimental}.
 A Gaussian distribution of 12000 $Sr^+$ ions and electrons a placed at the center of a spherical geometry that enclosed with a first order absorbing boundary condition. 
The initial temperature of $Sr^+$ ions is 1K and electrons are 100K placed at a  density  of  $5\times10^8$ particles  per  cubic  meter such that the particles are and will remain sufficiently away from the boundary. 
 Here, we use three geometries; the first with a radius of $6$ cm with an avergae edge length of 1.02 cm, the second $12$ cm with average edge length of 2.04 cm, and the third $18$ cm with average edge length of 3.06 cm.
 The experiment was run with first and second order basis functions, with a comparison to the analytic solution in Table \ref{tbl:table_of_figures}. Though there is good agreement between all of the experimental and analytic data, there is not a clear improvement as order increases. This is due to the fields being well behaved enough in this example that they are approximated well enough by the first order basis functions. 
     \begin{table}
     \addtolength{\tabcolsep}{-2pt}
        \centering
        \begin{tabular}{c|ccc}
           \toprule
            & $k$ & 1 & 2   \\
            \midrule
            \hline
            Radius & & &   \\
            6 cm & &\includegraphics[draft=false,scale=0.22]{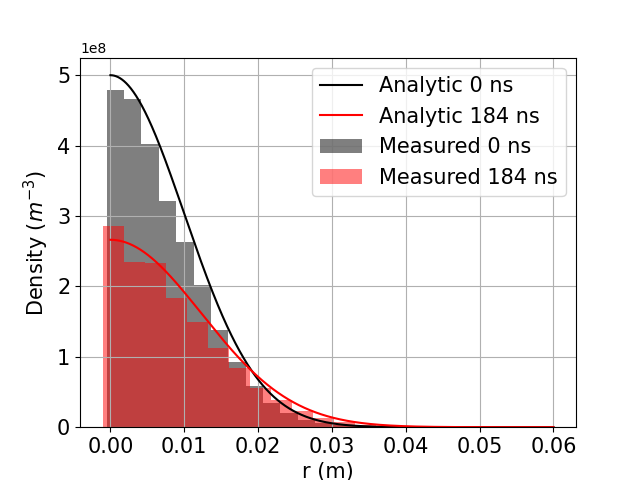} & \includegraphics[draft=false,scale=0.22]{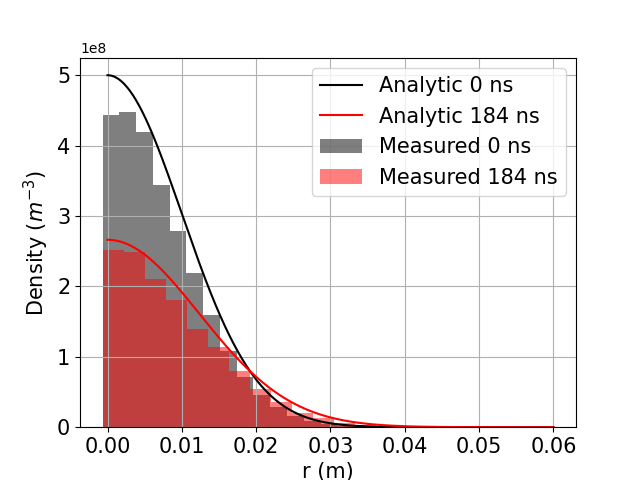} \\
            12 cm & & \includegraphics[draft=false,scale=0.22]{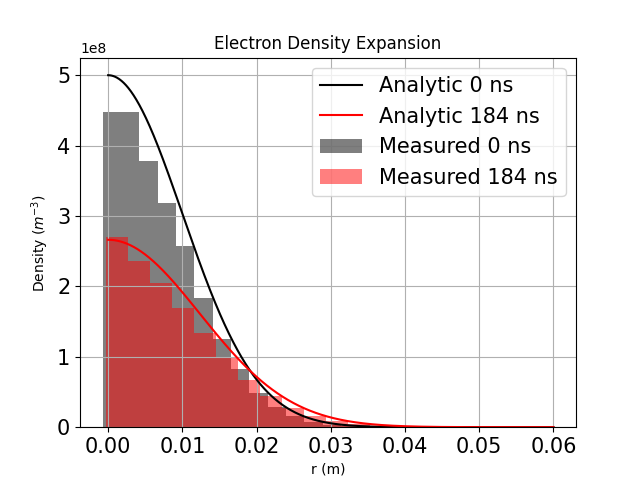} & \includegraphics[draft=false,scale=0.22]{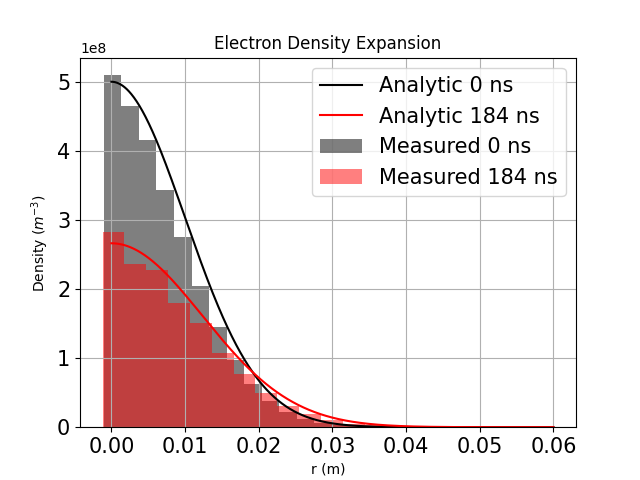}\\
            18 cm & & \includegraphics[draft=false,scale=0.22]{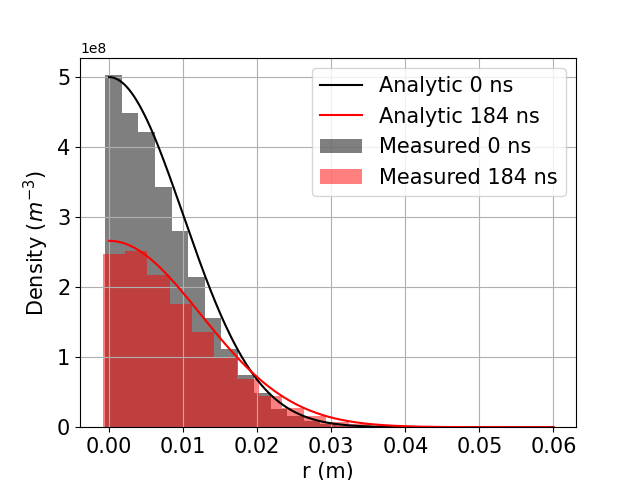} & \includegraphics[draft=false,scale=0.22]{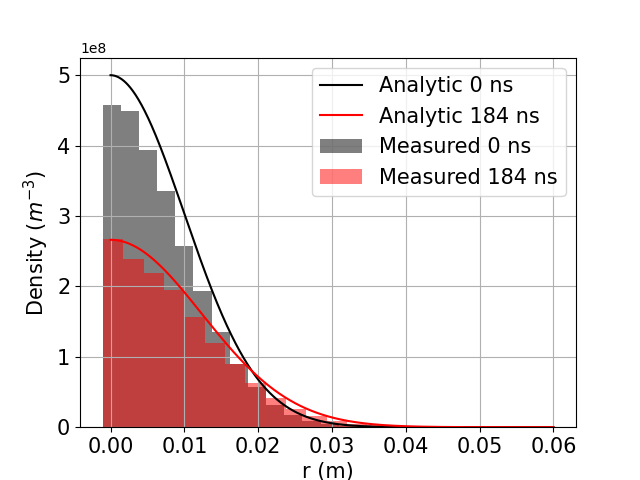}\\
            \bottomrule
        \end{tabular}
        \caption{Plasma Ball expansion for different basis function orders and geometry radii}
        \label{tbl:table_of_figures}
    \end{table}

\subsection{Expanding Particle Beam}

In this test, we demonstrate an expanding plasma beam in the PEC cavity.
This test is a standard test to confirm charge conservation as errors will accumulate and cause striations in the beam.
Additionally, quasi-analytic solutions and comparisons to this method and other discretization schemes can be found in \cite{reiser1994theory}. 
Macroparticles are injected into the cavity at an initial velocity and are allowed to repel each other as they progress down the cavity. 
The parameters used in this experiment are included in Table \ref{tb:beam}.
\begin{table}[ht!]
\centering
    \caption{Expanding Particle Beam Parameters}
    \begin{tabular}{c|c}
         \textit{Parameter} & \textit{Value}  \\
         \hline
         Cavity Radius & 20 mm \\
         Cavity Length & 100 mm\\
         Boundary Conditions & PEC \\
         $v_0$ & $1.02\cdot10^7$ m/s \\ 
         $v_0/c$ & 0.16678 \\ 
         beam radius $r_b$ & 8.00 mm \\
         Number particles per time step & 10 \\
         species & electrons \\
         Turn on time & 2 ns \\
         beam current  & 5 mA \\
         macro-particle size & 103921.12 \\
         min edge length & 3.89 mm \\
         max edge length & 13.5 mm \\
         $\Delta_t$ & 33.3 ps\\
    \end{tabular} \label{tb:beam}
\end{table}

\begin{figure}
    \centering
    \includegraphics[draft=false,scale=0.55]{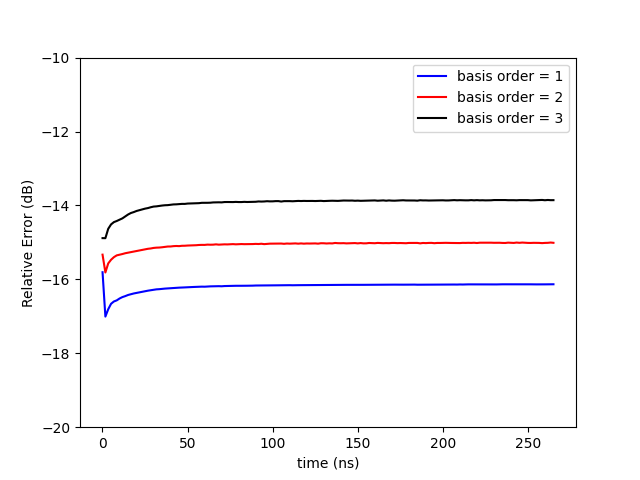}
    \caption{Relative Error of $\tilde{\rho}(t,\vb{r})$ vs $-[\grad]\vb{G}(t,\vb{r})$ for particle beam.}
    \label{fig:Grhobeamerr}
\end{figure}

\begin{figure}
    \centering
    \includegraphics[draft=false,scale=0.55]{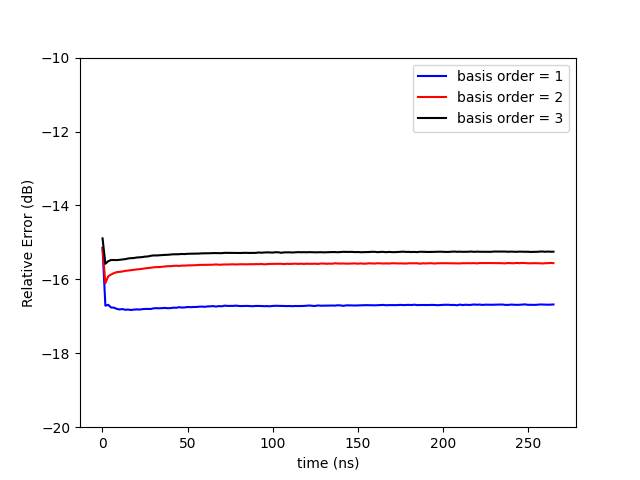}
    \caption{Relative Error of $[\div]\vb{D}(t,\vb{r})$ vs$-[\grad]\vb{G}(t,\vb{r})$ for particle beam.}
    \label{fig:divEbeamerr}
\end{figure}

\begin{figure}
    \centering
    \includegraphics[draft=false,scale=0.55]{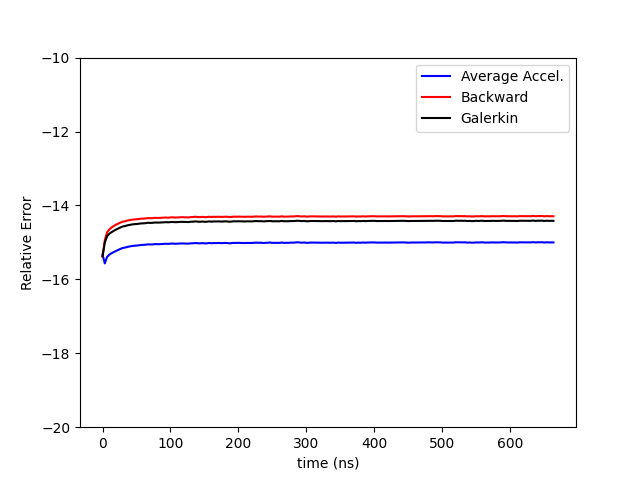}
    \caption{Relative Error of $[\div]\vb{D}(t,\vb{r})$ vs$-[\grad]\vb{G}(t,\vb{r})$ for particle beam with different time marching schemes.}
    \label{fig:alttserr}
\end{figure}

Key here is that the higher order basis functions are defined such that the differential relations between the basis functions for the electric field and magnetic flux density are preserved. 
First, consider the error in the measured charge density as defined by \eqref{eq:rerr} shown in Fig. \ref{fig:Grhobeamerr}. Here, the error is small and saturates as the total number of particles in the cavity stabilizes. 
The increase of error as the basis function order increases can be attributed to the increase of the condition number of the mass matrix that is inverted to compute the divergence of the integrated current.
The error between the divergence of the electric flux density and integrated current is shown in Fig. \ref{fig:divEbeamerr}.
The error is near machine precision, demonstrating the higher order divergence operator acting on the curl of the magnetic field does go to zero.
Lastly, different time marching schemes can be taken by varying the values of $\gamma$ and $\beta$ in \eqref{eq:maxnmb} using first order spatial basis functions.
The error in Gauss' law for three different time marching schemes is shown in \ref{fig:alttserr}, where average acceleration has $\gamma=.5, \beta=.25$, backward difference has $\gamma=1.5, \beta=1$,, and Galerkin has $\gamma=1.5, \beta=.8$. The error remains at machine precision as expected.
In total, this shows that the representation of the particle in the simulation is correct and that Gauss' law is satisfied.


\begin{table}
\addtolength{\tabcolsep}{-4pt}
\centering
\caption{Error in impressed Quadrupole Fields} \label{tab:quadErr}  
\begin{tabular}{|c c | c c | c c |}
\hline
  \multicolumn{2}{|c|}{basis order 1} &  \multicolumn{2}{|c|}{basis order 2} & \multicolumn{2}{|c|}{basis order 3}\\
\hline
 $N_2$ & error & $N_2$  & error  & $N_2$ & error \\
 \hline
 140629 & $9.835\times 10^{-2}$ & 89502 & $1.143\times 10^{-2}$ & 28488 & $1.726\times10^{-2}$ \\
 597623 & $3.8299\times 10^{-2}$ & 343812 & $3.803\times 10^{-3}$ & 103876 & $3.9481\times 10^{-3}$ \\
  &  & 637050 & $2.424\times 10^{-3}$ & 240270 & $6.932\times 10^{-4}$ \\
\hline
\end{tabular}
\end{table}

\subsection{Panofsky Quadrupole}

\begin{figure}
    \centering
    \includegraphics[draft=false,scale=0.35]{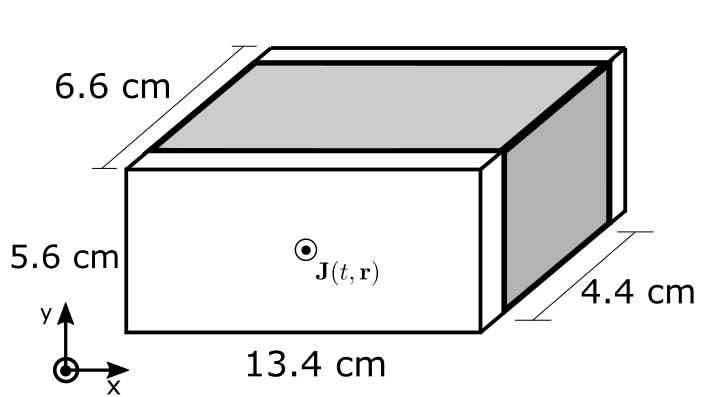}
    \caption{Schematic for perfect electrical conducting box with Pafonsky quadrapole (shaded). }
    \label{fig:quadSchem}
\end{figure}

\begin{figure}
    \centering
    \includegraphics[draft=false,scale=0.35]{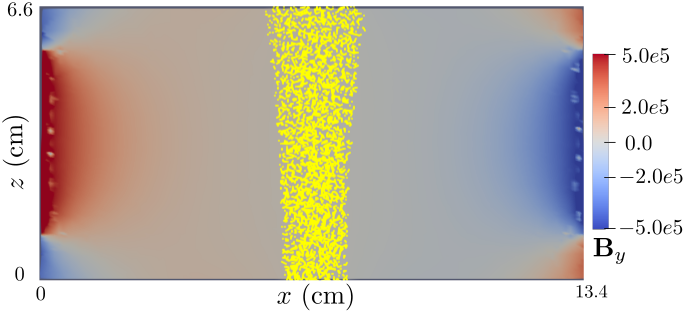}
    \caption{Particle beam traveling in $x-z$ plane in Pafonsky quadrapole with background $\vb{B}_y(t,\vb{r})$. }
    \label{fig:fldBx}
\end{figure}

\begin{figure}
    \centering
    \includegraphics[draft=false,scale=0.35]{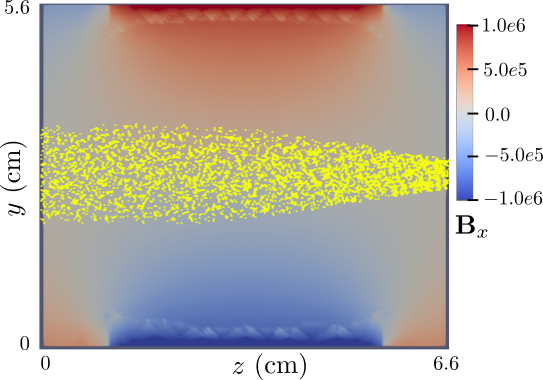}
    \caption{Particle beam traveling in $y-z$ plane in Pafonsky quadrapole with background $\vb{B}_x(t,\vb{r})$. }
    \label{fig:fldBy}
\end{figure}

In this example, the Panofsky quadrupole used in \cite{li1999design} is modeled using higher order basis to decrease the number of degrees of freedom. From Table \ref{tab:quadErr}, a mesh with more than several million degrees of freedom would be needed to obtain a result with similar error in the impressed fields. This would significantly effect the time necessary for both analysis and design.
The quadrupole is 13.4 cm in width, 5.6cm in height, and .044 cm in height, shown in Fig. \ref{fig:quadSchem}. We define a rectangular PEC cavity with the quadrupole .011cm from the xy-plane. The fields due to the quadrapole are calculated by evaluating Bio-Savart's law using the defined current density. It would be too expensive to recalculate the fields as particles pass through the system, therefore the coefficients needed to reconstruct the fields are obtained through \eqref{eq:conB}, which also enforces \eqref{eq:gaussB}. Particles are emitted from the xy-plane at $z=0$, passing through the fields generated by the quadrupole. 
The beam is emitted such that the beam will focus in the $y$-dimension and defocus in the $x$-dimension.
The iterative solver tolerance for both the fields and the divergence enforcement was set to 1e-5.
A snapshot of the particle positions at 35.7 ns is shown in Fig \ref{fig:fldBx} and \ref{fig:fldBy}, where the background magnetic flux density  $\vb{B}_y(t,\vb{r})$ and $\vb{B}_x(t,\vb{r})$ are plotted, respectively.
The particle beam smoothly expands in the $x$-dimension and compresses in the $y$-dimension which is a sign that charge is being conserved correctly.


\section{Summary \label{sec:conclusions}}

In this paper, we have presented a higher order, exact current mapped FEM-based EM-PIC with unconditionally stable time marching.
When higher order fields dominate the physics being modeled, higher order basis functions can reduce the number of degrees of freedom needed to model it while also getting more accurate fields.
A method was also provided to ensure that Gauss' magnetic law is satisfied, which can occur when time marching schemes other than leap frog are used.
Future work will create quasi-Helmholtz decomposition for higher order basis sets which will further increase the efficiency of using higher order basis while also allowing fields obtained by the vector wave equation to be used.
Even with the use of higher order basis, there is a limit to the size of problem that can fit on a single node.
To handle this challenge, domain decomposition approaches for Maxwell solvers will be developed.

\section{Acknowledgments}
This work was sponsored by the US Air Force Research Laboratory under contracts FA8650-19-F-1747 and FA8650-20-C-1132. We thank the MSU Foundation for support through the Strategic Partnership Grant during early portion of this work. This work was also supported by the SMART Scholarship program and the Department of Energy Computational Science Graduate Fellowship under grant DE-FG02-97ER25308. The authors would also like to thank the HPCC Facility,
Michigan State University, East Lansing, MI, USA.

\section{Appendix}

In this appendix, we define the higher order interpolatory basis functions. The basis functions are formed by multiplying the lowest order Whitney elements by an interpolatory Lagrange polynomial. We use the definition by Silvester \cite{silvester1996finite}
\begin{equation}
    P_i(\lambda) = \begin{cases}
    \frac{1}{i!}\prod_{m=0}^{i-1} (k\lambda - m), & 1\leq i\leq k\\
    1, & i=0
    \end{cases}
\end{equation}
and the shifted Silvester Polynomial
\begin{equation}
    \hat{P}_i(\lambda) = \begin{cases}
    \frac{1}{(i-1)!}\prod_{m=1}^{i-1} (k\lambda - m), & 2\leq i\leq k+1\\
    1, & i=1
    \end{cases}
\end{equation}

\subsection{Node Basis Function}
The k-th order nodal basis function $W_s^{(0)}(\vb{r}) \in H^1$ is defined as 
\begin{equation}
    W_i^{(0)}(\vb{r}) = Q^{(0)}\tilde{W}_i(\vb{r})
\end{equation}
where
\begin{equation}
    \tilde{W}_s^{(0)}(\vb{r}) = \lambda_s
\end{equation}
and
\begin{equation}
    Q^{(0)} = 
    \hat{P}_s(\lambda_s)P_t(\lambda_t)P_u(\lambda_u)P_v(\lambda_v).
\end{equation}
There are a total of $(k+1)(k+2)(k+3)/6$ degrees of freedom in each tetrahedron with $4$ associated with nodes, $k-1$ for each edge, $(k-1)(k-2)$ for each face, and $k(k-1)(k-2)/6$ internal to each tetrahedron.

\subsection{Whitney Edge Basis Function}
The k-th order edge basis function $\vb{W}_s^{(1)}(\vb{r}) \in H(\curl)$ is defined as
\begin{equation}
\vb{W}_s^{(1)}(\vb{r}) = Q^{(1)}\tilde{\vb{W}}^{(1)}_{st}(\vb{r})
\end{equation}
where 
\begin{equation}
    \tilde{\vb{W}}_s^{(1)}(\vb{r}) = \lambda_s\grad\lambda_t-\lambda_t\grad\lambda_s
\end{equation}
and
\begin{equation}
    Q^{(1)} = 
    \hat{P}_s(\lambda_s)\hat{P}_t(\lambda_t)P_u(\lambda_u)P_v(\lambda_v).
\end{equation}
There are a total of $k(k+2)(k+3)/2$ degrees of freedom in each tetrahedron with $k$ associated with each edge, $k(k-1)/2$ with each face, and $k(k-1)(k-2)/6$ internal to each tetrahedron.

\subsection{Whitney Face Basis Function}
The k-th order face basis function $\vb{W}_s^{(2)}(\vb{r}) \in H(\div)$ is defined as
\begin{equation}
\vb{W}_s^{(2)}(\vb{r}) = Q^{(2)}\tilde{\vb{W}}^{(2)}_s(\vb{r})
\end{equation}
where 
\begin{equation}
    \tilde{\vb{W}}_s^{(2)}(\vb{r}) = \lambda_s\grad\lambda_t\cross\grad\lambda_u+\lambda_t\grad\lambda_u\cross\grad\lambda_v+\lambda_u\grad\lambda_s\cross\grad\lambda_t
\end{equation}
and
\begin{equation}
    Q^{(2)} = 
    \hat{P}_s(\lambda_s)\hat{P}_t(\lambda_t)\hat{P}_u(\lambda_u)P_v(\lambda_v).
\end{equation}
There are a total of $k(k+1)(k+3)/2$ degrees of freedom in each tetrahedron with $k(k+1)/2$ with each face and $k(k-1)(k+1)/6$ internal to each tetrahedron.

\subsection{Volume Basis Function}

The k-th order volume basis function $\tilde{W}^{(3)}(\vb{r}) \in L^2$ is defined as
\begin{equation}
\begin{split}
    W_s^{(3)}(\vb{r}) &= Q^{(3)}\tilde{W}^{(3)}(\vb{r})
\end{split}
\end{equation}
where 
\begin{equation}
\begin{split}
    \tilde{W}^{(3)}(\vb{r}) &= \lambda_s\grad\lambda_t\cdot\grad\lambda_u\cross\lambda_v +\lambda_t\grad\lambda_u\cdot\grad\lambda_v\cross\lambda_s\\
    &+\lambda_u\grad\lambda_v\cdot\grad\lambda_s\cross\lambda_t +\lambda_v\grad\lambda_s\cdot\grad\lambda_t\cross\lambda_u
\end{split}
\end{equation}
and
\begin{equation}
    Q^{(3)} = 
    \hat{P}_s(\lambda_s)\hat{P}_t(\lambda_t)\hat{P}_u(\lambda_u)\hat{P}_v(\lambda_v).
\end{equation}
The function is associated with tetrahedra with $k(k+1)(k+2)/6$ unknowns in each tetrahedron, all of which are internal to the cell.
 \bibliographystyle{IEEEtran}
 \bibliography{main}

\end{document}